# Kidney and Kidney Tumour Segmentation in CT Images

Qi Ming How[1, a)] and Hoi Leong Lee[2, b)]

[1]*Faculty of Electronic Engineering Technology, Universiti Malaysia Perlis, Arau, 02600 Malaysia*

a) Corresponding author: dannyhow98@gmail.com
b) hoileong@unimap.edu.my

**Abstract.** Automatic segmentation of kidney and kidney tumour in Computed Tomography (CT) images is essential, as it uses less time as compared to the current gold standard of manual segmentation. However, many hospitals are still reliant on manual study and segmentation of CT images by medical practitioners because of its higher accuracy. Thus, this study focuses on the development of an approach for automatic kidney and kidney tumour segmentation in contrast-enhanced CT images. A method based on Convolutional Neural Network (CNN) was proposed, where a 3D U-Net segmentation model was developed and trained to delineate the kidney and kidney tumour from CT scans. Each CT image was pre-processed before inputting to the CNN, and the effect of down-sampled and patch-wise input images on the model performance was analysed. The proposed method was evaluated on the publicly available 2021 Kidney and Kidney Tumour Segmentation Challenge (KiTS21) dataset. The method with the best performing model recorded an average training Dice score of 0.6129, with the kidney and kidney tumour Dice scores of 0.7923 and 0.4344, respectively. For testing, the model obtained a kidney Dice score of 0.8034, and a kidney tumour Dice score of 0.4713, with an average Dice score of 0.6374.

## INTRODUCTION

Kidneys function in the urinary system to filter blood, maintain fluid balance and electrolyte levels, and to produce hormone erythropoietin which stimulates stem cells to produce red blood cells. In line with this, kidney tumours have been posing as a serious health issue, with its detection aided by medical imaging modalities like Magnetic Resonance Imaging (MRI), X-Ray and Computed Tomography (CT). These assists in diagnosis, treatment planning and clinical research for medical practitioners. A more refined method is devised, which is medical image segmentation. This intends to delineate the Region of Interest (ROI) of the CT scans, for example segmenting the region of kidney and kidney tumour automatically. This is aided by the fast-developing computer vision algorithms, in addition to the boom of deep learning in the past decade, most notably in the success of ImageNet classification in 2012 [1] and the introduction of U-Net in 2015 [2]. In line with this, Kidney and Kidney Tumour Segmentation Challenge 2019 (KiTS19) and 2021 (KiTS21) [3] are introduced to encourage progress in the field of medical image analysis and automatic segmentation. However, many hospitals are still reliant on the process of manual segmentation of organs and lesions of interest done by medical practitioners, and the impact of the challenges have not been seen in hospital settings. The issue with this is that manual segmentation of CT scans is a time-consuming process, as repetitive effort from medical practitioners to produce high-quality 3D segmentation of the kidney and kidney tumours is required. Also, many methods developed for automatic kidney and kidney tumour segmentation are not validated, thus could mean that the method developed will not perform well in a new set of data.

In line with this, participants of KiTS19 achieved considerable success, shown in the work done by [4], where a Dice score of 0.974 for kidney and Dice score of 0.851 for kidney tumour are obtained. Their work proposed to train three different CNN models which are the plain 3D U-Net, residual 3D U-Net and pre-activation 3D U-Net, with the model trained with residual 3D U-Net recording the best score. In addition, several findings have also performed pre-processing before inputting the images into the CNN model [3]. In the study done by [5], several loss functions are proposed to handle class imbalance dataset, where it is concluded that region-based loss function such as Dice loss function perform better with class imbalanced data. Also, cross-entropy loss function is a distribution-based loss function where it provides a suitable measure to distinguish between two or more discrete classes. At the end of their study, the model with Dice loss function performed better as compared to the model with cross-entropy loss function, with the best performing model recorded a kidney Dice score of 0.931 and kidney tumour Dice score of 0.536.

Therefore, this study may be essential for the development of a reliable method for kidney and kidney tumour segmentation in CT scans. The contributions targeted from this study is as follows:
- To develop an approach based on Convolutional Neural Network (CNN) for automatic kidney and kidney tumour segmentation in CT scan images.
- To evaluate the effect of down-sampled and patch-wise input images on the performance of CNN model.
- To validate the developed method on the dataset of 2021 Kidney and Kidney Tumour Segmentation Challenge.

This paper is organised as follows: Section 2 presents the preparation of data and pre-processing techniques performed. Section 3 highlights the implementation details of the proposed method. Section 4 explains the results and discussion obtained. Lastly, Section 5 concludes this study.

## DATA PREPARATION AND PRE-PROCESSING

Throughout this study, the proposed method is developed with Python programming language, and the Medical Open Network for Artificial Intelligence (MONAI) framework, which is an open-source foundation created by Project MONAI [6]. It is built based on the PyTorch machine learning framework, and the proposed algorithm is debugged through the web-based Integrated Development Environment (IDE), Google Colaboratory with Graphical Processing Unit (GPU) implementation. The dataset used in this study is contrast-enhanced CT images in the format of Neuroimaging Informatics Technology Initiative (NIfTI). 300 CT scans and its corresponding segmentation masks are made available by KiTS21 organisers, where only 125 CT scans are used throughout this study to ease memory consumption. Out of the 125 images, it is split into 80:20 ratio, where 100 images are used as the training dataset while 25 images are used as the testing dataset. Upon analysis, each CT image are of the size 512 x 512 x z voxels, where z varies between 34 to 834 for all 125 images. Therefore, several pre-processing techniques are used to standardise the size of the input image to the CNN model.

The pre-processing methods used are based on the MONAI framework, and is automated with the algorithm written. Several pre-processing techniques are used, which are unification of voxel spacing, scaling of the intensity of CT image, foreground cropping, down-sampling and random cropping. Upon analysis, all 125 CT scans have varying voxel spacing, where the x and y-axis ranges from 0.537 – 0.977, and z-axis ranges from 0.5-5. Due to this, the voxel spacing is standardised to a value of 1.62 x 1.62 x 3.22, where this value is obtained in reference of the success achieved by [4]. Then, the intensity of the images is clipped in the range of [-79, 304], where this is a range of organ specific value for kidney [4]. Then, foreground cropping is performed, where this effectively crops the foreground region of the image. This is because the CT scans have borders in the CT image that does not represent the ROI in this study, which is the region of kidney and kidney tumour. The foreground region is determined by the source image where this performs a search for the valid foreground region with reference to the raw image, therefore removing any additional background information, effectively narrowing down the Field of View (FOV).

For the training of the first model, the images are down-sampled to 128 x 128 x 32 before going into the CNN model. Down-sampling is done by trilinear interpolation, an extension of bilinear interpolation, where it is the linear interpolation of two bilinear interpolations. This is normally done to a volumetric size, where in this study to down-sample the volumetric size of the 3D CT image. So far, most methods developed are to extract patches with a fixed protocol, with limited analysis on the effect of random cropping on performance of the CNN model. In this study, random cropping is performed for the training of the second model, where 4 crops are randomly generated for each CT scan. In the third model, the input images will first be down-sampled to 128 x 128 x 32, then with 4 random crops extracted from each down-sampled image. Table 1 summarises the pre-processing techniques performed for the 3 models, while Equation 1 explains the acquisition of crops in random cropping.

$$Ratio = \frac{pos}{pos+neg} \quad (1)$$

Pos and neg is set to 1 respectively, to ensure that there is always a 50% chance to select a crop with a foreground voxel as the centre of the crop.

TABLE 1. Pre-processing for 3 models.

| Model 1 | Model 2 | Model 3 |
|---|---|---|
| | Unification of voxel spacing to 1.62 x 1.62 x 3.22 | |
| | Scaling of intensity in the range of [-79, 304] | |
| | Foreground cropping | |
| Down-sampling to 128 x 128 x 32 | Random cropping of 4 samples with size 128 x 128 x 32 | Down-sampling to 128 x 128 x 32, then random cropping of 4 samples with size 128 x 128 x 32 |

## IMPLEMENTATION DETAILS

The CNN model that will be used in this study is the 3D U-Net, where the 3D volumes which have undergone pre-processing methods mentioned above will be input into this model. This can be visualised in Figure 1.

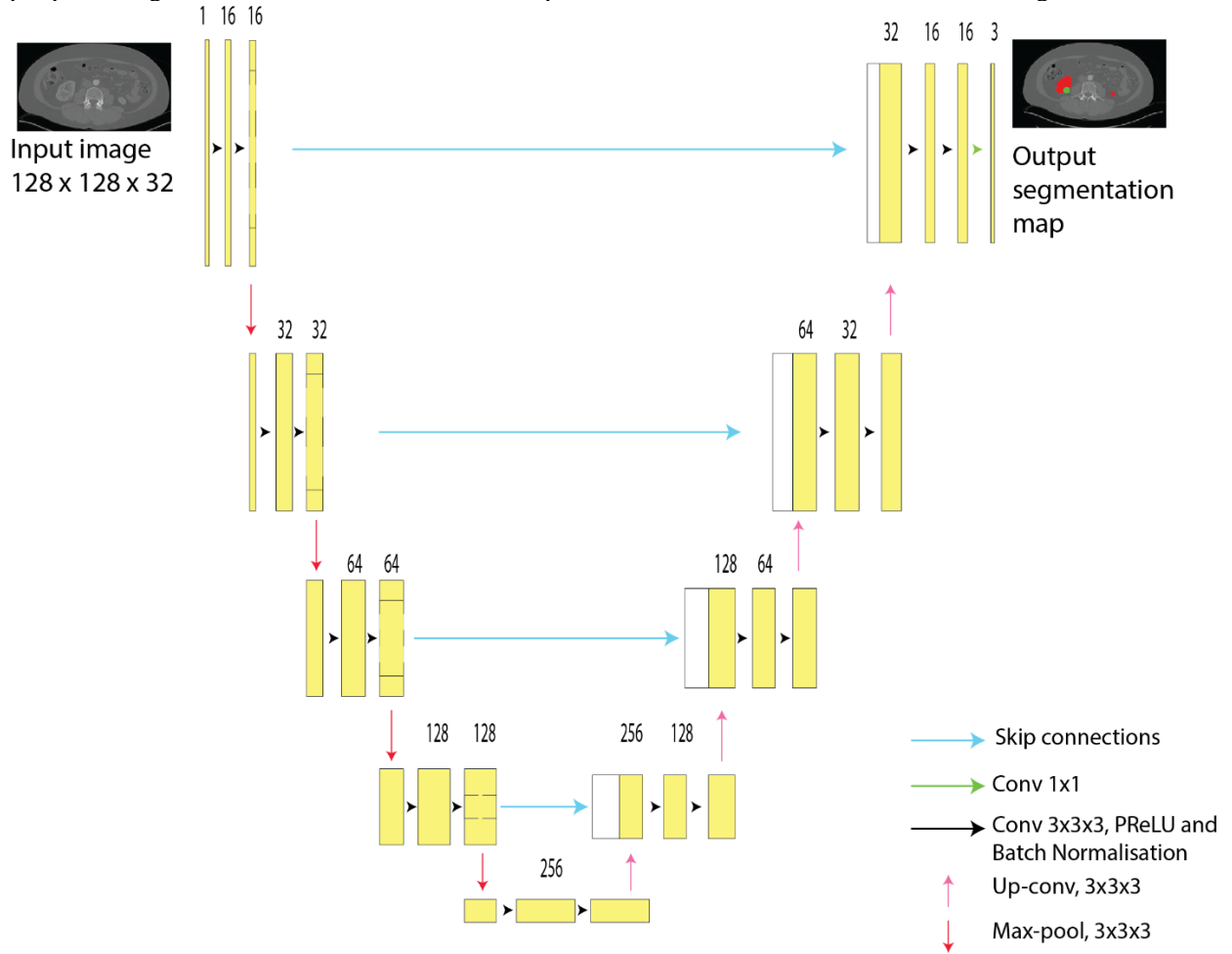

**FIGURE 1.** Proposed U-Net architecture for training of the segmentation model

From Figure 1, it is observed that the input images are of the size 128 x 128 x 32, and the output segmentation map will consist of 3 classes, which are the background, kidney and kidney tumour. In addition, residual unit is used in the down-sampling layer of the encoding path, while also applying batch normalisation after each convolution layer. Parametric Rectified Linear Unit (PReLU) is used as the activation function, explained with Equation 2.

$$f(x_i) = \begin{cases} x, & if\ x_i \geq 0 \\ a_i x_i, & if\ x_i < 0 \end{cases} \quad (2)$$

$a_i$ is a learnable parameter, where in comparison with Rectified Linear Unit, the value of $a_i = 0$. Also, the loss function that is used in this study is the Dice loss function, where it is obtained by the formula of Dice loss = 1 – Dice, where Dice is shown in Equation 3.

$$Dice = \frac{2\,|A \cap B| + smooth}{(|A| + |B| + smooth)} \quad (3)$$

A is the predicted value of kidney and B is the ground truth value of kidney for the kidney class, similarly for the kidney tumour class where A is the predicted value of kidney tumour and B is the ground truth value of the kidney tumour class. Smooth is generally set to 1, where this is added to prevent the division by 0 during a scenario when A = B, leading to an undefined Dice [7]. Table 2 summarises the implementation details of the training process. During the computation of Dice, the background class will be neglected, due to the size of the kidney and kidney tumour being smaller as compared to the background. This can be visualised in Figure 2.

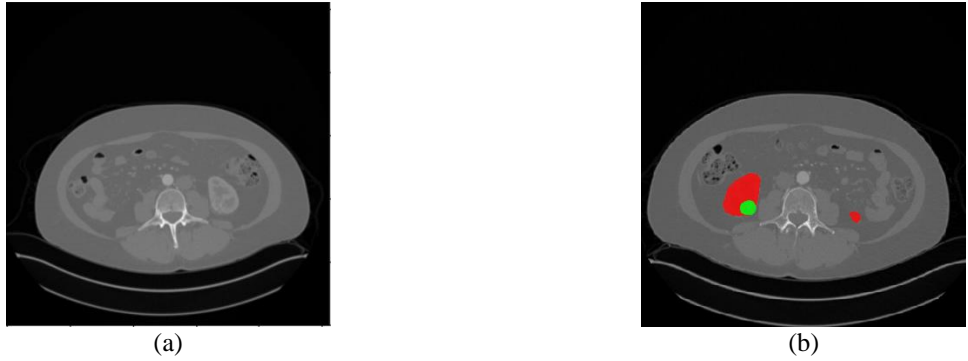

(a)          (b)

**FIGURE 2**. Example of contrast-enhanced CT image and the overlay of segmentation mask on the CT image.

**TABLE 2.** Implementation details.

| Parameter | Value/Method |
|---|---|
| Activation function | PReLU |
| Activation function (output layer) | Softmax |
| Loss function | Dice |
| Optimiser | Adam |
| Learning rate | $1e^{-4}$ |
| Learning rate scheduler | Reduce by factor of 10 |
| Epochs | 300 |
| Batch Size | 1 |

From Table 2, it is observed that learning rate scheduler will be implemented, where this monitors the validation loss. If the validation loss does not reduce in 10 consecutive epochs, then the learning rate will be reduced by a factor of 10. Each model will be trained for 300 epochs with a learning rate of $1e^{-4}$ and a batch size of 1, and it will be performed with GPU provided by Google Colaboratory.

## RESULTS AND DISCUSSION

After training of the 3 CNN models, it is then tested against the test dataset to validate the performance of the CNN model. The total time required to train model 1 across all 300 epochs is 6889.62s, as compared to 15238.45s and 12727.07s for model 2 and model 3 respectively. This is because for model 2 and model 3, random cropping is performed. Due to the method of CacheDataset made available by MONAI, random transforms are not cached and will be perform before each training of the batch. Therefore, this causes an increase in training time for model 2 and model 3.

**TABLE 3.** Comparison of testing Dice score for the three models

| Dice | Model 1 | Model 2 | Model 3 |
|---|---|---|---|
| Average | 0.0356 | 0.6374 | 0.0400 |
| Kidney | 0.0427 | 0.8034 | 0.0424 |
| Tumour | 0.0284 | 0.4713 | 0.0376 |

In Table 3, it is observed that model 2 is best performing compared to model 3 as it recorded the highest kidney Dice score and kidney tumour Dice score, which is followed by model 1 which performs the worst. The reason why model 1 and model 3 performed poorly could be due to the loss of image information as the images were down-sampled. In addition, model 3 perform slightly better than model 1, and it may be due to more samples being used during the training process.

**TABLE 4.** Performance analysis of best model on test dataset.

| Case | Kidney | Kidney Tumour Dice | Average Dice | Time (s) |
|---|---|---|---|---|
| 36 | 0.8038 | 0.7103 | 0.7571 | 24.24 |
| 37 | 0.4276 | 0.6835 | 0.5556 | 15.79 |
| 39 | 0.8002 | 0 | 0.4001 | 16.10 |
| 41 | 0.9341 | 0 | 0.4671 | 8.91 |
| 42 | 0.8872 | 0.1160 | 0.5016 | 39.90 |
| 43 | 0.9152 | 0.2427 | 0.5790 | 26.54 |
| 44 | 0.9148 | 0.8552 | 0.8850 | 19.59 |
| 45 | 0.8855 | 0.6735 | 0.7795 | 10.66 |
| 46 | 0.9131 | 0.8909 | 0.9020 | 24.40 |
| 47 | 0.7889 | 0.7638 | 0.7764 | 24.32 |
| 48 | 0.9653 | 0.8296 | 0.8975 | 17.62 |
| 50 | 0.9662 | 0 | 0.4831 | 15.33 |
| 51 | 0.9543 | 0.7102 | 0.8323 | 11.27 |
| 54 | 0.7086 | 0.1444 | 0.4265 | 21.27 |
| 55 | 0.7780 | 0.8477 | 0.8129 | 17.07 |
| 56 | 0.4346 | 0.7855 | 0.6101 | 15.40 |
| 57 | 0.8188 | 0.7280 | 0.7734 | 13.43 |
| 58 | 0.7549 | 0.4318 | 0.5934 | 18.11 |
| 60 | 0.8374 | 0.7843 | 0.8109 | 22.15 |
| 61 | 0.2921 | 0.2175 | 0.2548 | 8.03 |
| 62 | 0.8073 | 0.0320 | 0.4197 | 16.49 |
| 64 | 0.9532 | 0.0117 | 0.4825 | 9.54 |
| 65 | 0.9295 | 0.3858 | 0.6577 | 16.01 |
| 69 | 0.8474 | 0.1442 | 0.4958 | 15.20 |
| 70 | 0.7679 | 0.7944 | 0.7812 | 10.61 |

Table 4 tabulates all kidney and kidney tumour Dice of the 25 images for model 2. It is observed that there are a few cases which recorded a tumour Dice score of 0, which are case 39, 41 and 50. Also, the average time required to generate the segmentation mask for each case is 17.52 seconds with GPU implementation. On the other hand, Table 5 shows the comparison of result for training and testing for model 2. The training Dice score and testing Dice score is almost similar, thus proving that the model is not overfitting.

**TABLE 5.** Comparison of Dice score for training and testing of best model.

| Dice | Training | Testing |
|---|---|---|
| Average | 0.6129 | 0.6374 |
| Kidney | 0.7923 | 0.8034 |
| Tumour | 0.4344 | 0.4713 |

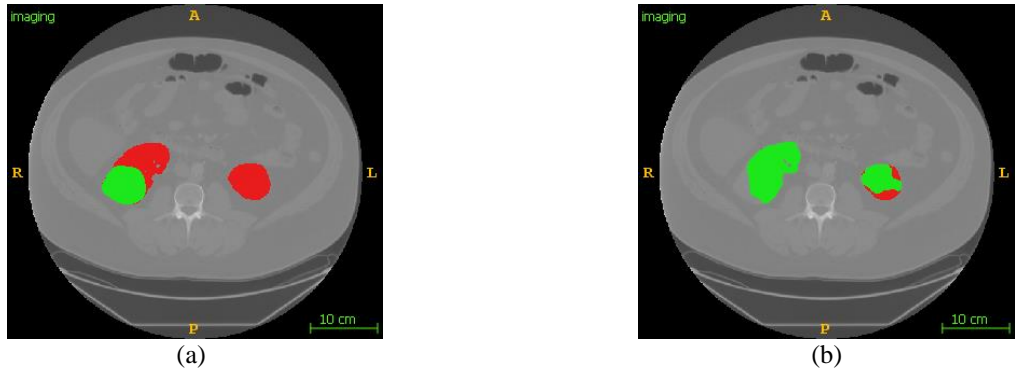

(a)          (b)

**FIGURE 3**. Axial view of ground truth mask and segmentation mask generated for case 61. On the segmentation mask generated, it is observed that the whole kidney is predicted to be the region of the tumour, represented by the green coloured region. The kidney Dice and kidney tumour Dice computed is 0.2921 and 0.2175 respectively.

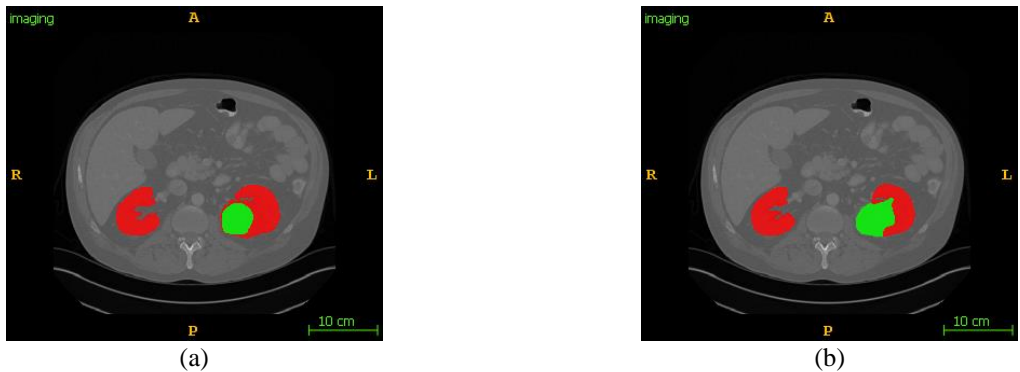

(a)          (b)

**FIGURE 4**. Axial view of ground truth mask and segmentation mask generated for case 36. This shows a more accurate annotation of the kidney region, however there are parts of the kidney that is predicted to be kidney tumour. The kidney Dice and kidney tumour Dice computed is 0.8038 and 0.7103 respectively.

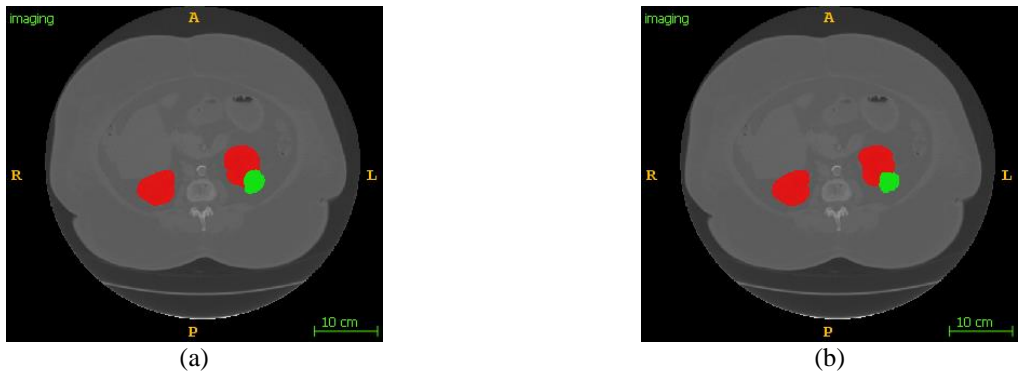

**FIGURE 5**. Axial view of ground truth mask and segmentation mask generated for case 48. The segmentation mask predicted shows a close to perfect prediction of both the region for kidney and kidney tumour. The kidney Dice and kidney tumour Dice computed is 0.9653 and 0.8296 respectively.

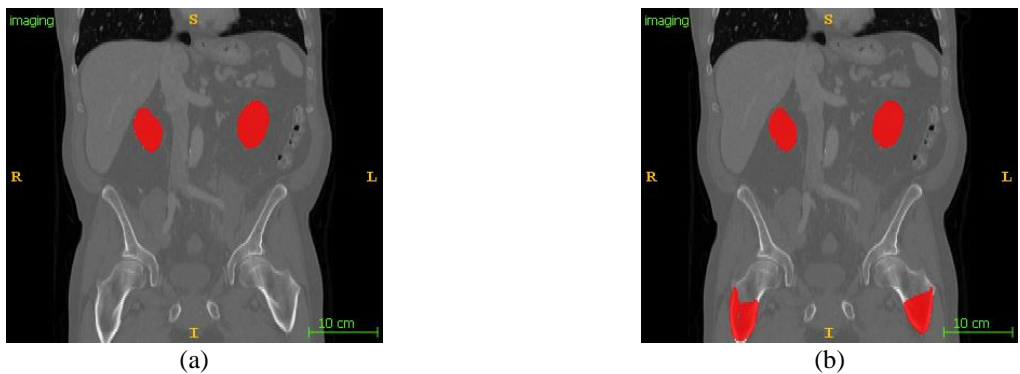

**FIGURE 6**. Coronal view of ground truth mas and segmentation mask generated for case 36. As observed in the segmentation mask generated, the two red regions at the bottom of the mask are predicted to be kidney, which is not the case as shown in the ground truth mask.

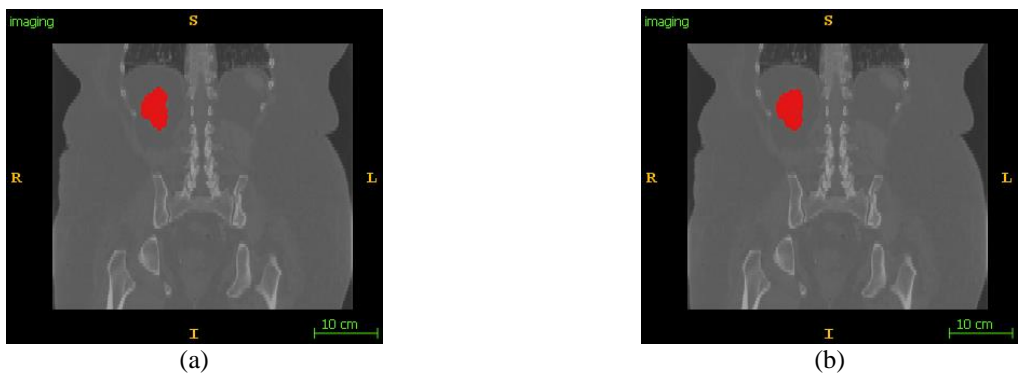

**FIGURE 7**. Coronal view of ground truth mask and segmentation mask generated for case 48. Similarly, the segmentation mask generated on case 48 is close to perfect as shown in Figure 5, even from a different view.

## CONCLUSION

Due to the long time required to perform manual segmentation, an automatic method for kidney and kidney tumour segmentation is proposed. A CNN based approach has been developed to perform segmentation of kidney and kidney tumour from contrast-enhanced CT scans.

Several pre-processing methods performed on the training data include unification of voxel spacing, scaling of intensity, and foreground cropping. Three separate models were then trained with down-sampled, patch-wise input image and the combination of these two techniques respectively. Among the three models, the best performing was model 2 which recorded a kidney Dice score of 0.8034 and kidney tumour Dice score of 0.4713.

## ACKNOWLEDGEMENTS

I would like to thank my supervisor, Dr Lee Hoi Leong, for his guidance and support in making this research possible.